\documentclass[aps,prl,superscriptaddress,showpacs,twocolumn,10pt]{revtex4-1}

\usepackage{graphicx}
\usepackage{amsmath}
\usepackage{color}
\usepackage[colorlinks=true,linkcolor=blue,urlcolor=blue,citecolor=blue]{hyperref}
\usepackage{ulem}
\usepackage{mathrsfs}
\usepackage{xspace}
\usepackage[utf8]{inputenc}

\newcommand{\ket}[1]{\ensuremath{\left| #1 \right\rangle}\xspace} 
\newcommand{\up}{\ensuremath{\ket{\uparrow}}\xspace}
\newcommand{\down}{\ensuremath{\ket{\downarrow}}\xspace}

\begin{document}

\title{Spatially Resolved Detection of a Spin-Entanglement Wave in a Bose-Hubbard Chain}

\author{Takeshi Fukuhara}
\thanks{T.F. and S.H. contributed equally to this work.}
\email[Corresponding author: ]{takeshi.fukuhara@riken.jp}
\affiliation{Max-Planck-Institut f\"{u}r Quantenoptik, 85748 Garching, Germany}
\affiliation{RIKEN Center for Emergent Matter Science (CEMS), Wako, 351-0198, Japan}
\author{Sebastian Hild}
\thanks{T.F. and S.H. contributed equally to this work.}
\author{Johannes Zeiher}
\author{Peter Schauß}
\affiliation{Max-Planck-Institut f\"{u}r Quantenoptik, 85748 Garching, Germany}
\author{Immanuel Bloch}%
\affiliation{Max-Planck-Institut f\"{u}r Quantenoptik, 85748 Garching, Germany}
\affiliation{Ludwig-Maximilians-Universit\"{a}t, Fakult\"{a}t f\"{u}r Physik, 80799 M\"{u}nchen, Germany}%
\author{Manuel Endres}
\affiliation{Department of Physics, Harvard University, Cambridge, MA 02138, USA}
\author{Christian Gross}%
\affiliation{Max-Planck-Institut f\"{u}r Quantenoptik, 85748 Garching, Germany}

\date{\today}

\begin{abstract}
Entanglement is an essential property of quantum many-body systems.
However, its local detection is challenging and was so far limited to spin degrees of freedom in ion chains.
Here we measure entanglement between the spins of atoms located on two lattice sites in a one-dimensional Bose-Hubbard chain which features both local spin- and particle-number fluctuations.
Starting with an initially localized spin impurity, we observe an outwards propagating entanglement wave and show quantitatively how entanglement in the spin sector rapidly decreases with increasing particle-number fluctuations in the chain.
\end{abstract}


\maketitle
Quantum many-body systems are distinct from their classical counterparts due to entanglement among their constituents~\cite{Amico:2008, Eisert:2010}.
Especially in strongly correlated regimes, such as, in the vicinity of quantum phase transitions~\cite{Osterloh:2002, Amico:2008} or far away from equilibrium~\cite{Calabrese:2009}, the growth of entanglement with time or subsystem size seriously limits numerical simulations of complex quantum systems~\cite{Eisert:2010}.
Next to its importance on this fundamental level, entanglement is a valuable resource for quantum information and its microscopic control is required for most applications~\cite{Bennett:2000}.
Experimentally, the measurement of entanglement is difficult given that full quantum state tomography requires extraordinary control and resources making it feasible only in small systems~\cite{Roos:2004, Haffner:2005, Jurcevic:2014}.
In larger or more complex many-body systems the, mere presence of entanglement can be inferred from macroscopic observables, often relying on entanglement witnesses~\cite{Amico:2008, Guehne:2009}.
Such a strategy has been applied for susceptibility measurements in solids~\cite{Brukner:2006}, collective spin systems~\cite{Esteve:2008, SchleierSmith:2010, Gross:2010, Riedel:2010, Gross:2011, Haas:2014} or coupled superconducting qubits~\cite{Lanting:2014}.

In Hubbard systems realized with ultracold atoms, entanglement in the on-site occupation number degree of freedom has been inferred from the visibility of a far-field interference pattern~\cite{Cramer:2011, Cramer:2013}.
A spatially resolved detection of entanglement has been recently proposed using the R\'{e}nyi entropy~\cite{Daley:2012, Abanin:2012}.
Extension of the Hubbard model to two components introduces a spin degree of freedom~\cite{Kuklov:2003, Duan:2003} such that spin-entanglement, in the sense of the concept of entanglement of particles~\cite{Wiseman:2003, Dowling:2006}, can be present.
First experiments with ultracold atoms showed that short-range coherent spin-dynamics can be controlled in bosonic~\cite{Trotzky:2008, Trotzky:2010, Nascimbene:2012}, as well as in fermionic systems~\cite{Greif:2013}.
Spin exchange collisions in state selective optical lattices have been used to realize collisional gates between neighboring atoms~\cite{Jaksch:1999} and global measurements indicated entanglement~\cite{Mandel:2003, Anderlini:2007}.
However, a spatially resolved detection of either spin or occupation number entanglement in Hubbard models has still been an outstanding experimental challenge.

Inspired by recent measurements in ion chains~\cite{Jurcevic:2014}, here we report on the spatially resolved detection of entanglement among spin degrees of freedom in a two component Bose-Hubbard chain following a recent proposal~\cite{Mazza:2015}.
Specifically, we used local detection~\cite{Sherson:2010, Bakr:2009} and manipulation~\cite{Weitenberg:2011} to study the dynamics of a single spin-impurity.
Measurements in the longitudinal basis revealed the position of the impurity~\cite{Fukuhara:2013a, Fukuhara:2013b}, while transverse correlation measurements were used to extract its coherence.
Combining both observables yields a lower bound for the concurrence~\cite{Wooters:1998, Mintert:2004} in the spin-degrees of freedom of particles on two lattice sites~\cite{Mazza:2015}.
We observe an outward propagating entanglement wave, the evolution of which we follow up to a distance of six lattice sites.
Importantly, the detected bound is valid despite on-site particle-number fluctuations and proves entanglement of particles~\cite{Wiseman:2003, Schuch:2004, Dowling:2006} in our system~\cite{Mazza:2015}.
To study the effect of occupation-number defects on the entanglement propagation in more detail, we developed a novel in-situ Stern-Gerlach imaging technique that yields information on both, the local spin and the occupation-number in one image.

\begin{figure}[t]
\centering
\includegraphics[]{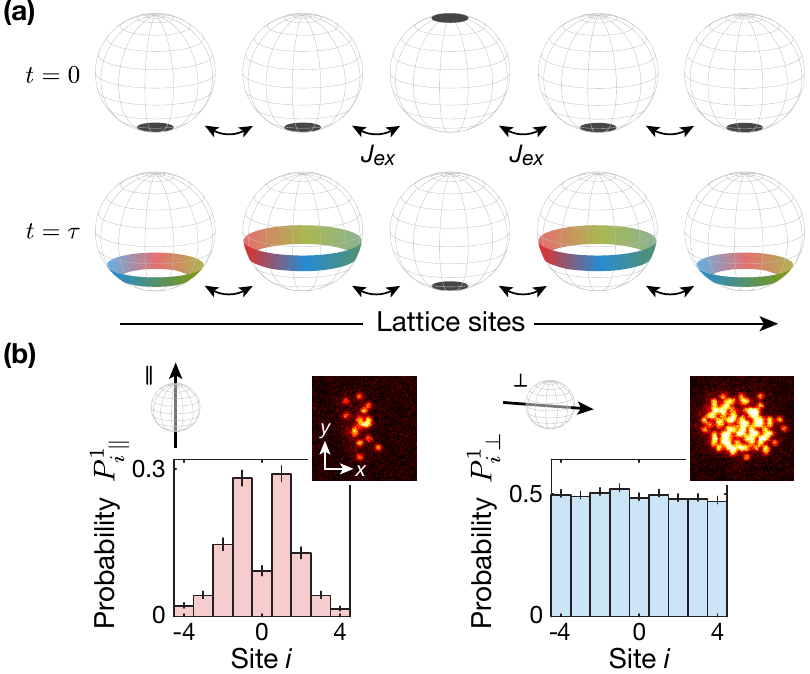}
\caption{Single spin-impurity dynamics.
(a) Schematics. The Bloch spheres represent the spin vector at different sites along the chain; their coupling with strength $J_{ex}$ is indicated by the arrows.
The $\up$ ($\down$) states are shown as the gray areas at the north (south) poles.
At time $t=0$, a single spin-impurity is deterministically created in the center and subsequently propagates to both, left and right.
For all times, the only non-vanishing first moment is $\langle \hat{S}^z_i \rangle$, the longitudinal mean spin.
The transverse direction is fully undefined (depicted for some later time $t=\tau$ by the ring on the Bloch spheres), but the dynamics generates correlations that are represented by the changing color along the ring, where the same color indicates correlations in the spin direction between the sites.
The size of the gray areas and the width of the ring are optimized for visibility and they have no physical meaning.
(b) Representative experimental measurements in the longitudinal (left) and transverse basis (right) after $t=35\,$ms.
The Bloch sphere pictograms indicate the measurement direction.
The images show exemplary single shot images.
The histograms show the probability at each lattice site to find an atom in the respective spin-resolved measurement.}
\label{fig:schematics}
\end{figure}

In our experiment, we realized ferromagnetic Heisenberg spin chains with two-component ultracold bosonic atoms in an optical lattice~\cite{Kuklov:2003, Duan:2003}.
In the unity filling Mott insulator regime and for equal inter- and intra-component scattering lengths, the system is described by the isotropic Heisenberg model $\hat{H}=-J_{ex}\sum_{i}(\hat{S}_i^x \hat{S}_{i+1}^x + \hat{S}_i^y \hat{S}_{i+1}^y + \hat{S}_i^z \hat{S}_{i+1}^z) + \hat{H}_d$, where the last term accounts for defects (holes or multiple occupancies) in the chain.
The operators $\hat{S}_i^\alpha$ denote the components of a spin-$1/2$ operator at site $i$ with $\alpha=x,y,z$ and $J_{ex} \approx 4J^2/U$  is the superexchange coupling for onsite interaction $U$ and tunnel coupling $J$ of the underlying Bose-Hubbard model.
We prepared the system in a fully polarized state (i.e. the spin of all atoms points down) and created a single spin-up impurity on the central site.
In this case, the interaction term $\hat{S}_i^z \hat{S}_{i+1}^z$ can be dropped and the Hamiltonian reduces to the XX-model in the spin sector.
The wavepacket of the spin-impurity dispersed with evolving time resulting in a build up and subsequent spreading of spin-entanglement along the chain~\cite{Bose:2003, Subrahmanyam:2004, Amico:2004, Mazza:2015}.
As an experimentally accessible entanglement measure, we use the concurrence between two lattice sites~\cite{Wooters:1998,Mintert:2004}, which measures entanglement on a scale between zero (no entanglement) and one (maximal entanglement).
A convenient lower bound, detectable with only global spin rotations, is given by $\mathscr{C}_{i,j} = 2 (2C_{i,j} - (P^{\uparrow, \uparrow}_{i,j} P^{\downarrow,\downarrow}_{i,j})^{1/2})$~\cite{Mazza:2015}.
The first term $C_{i,j} = \langle \hat{S}_i^\perp \hat{S}_{j}^\perp \rangle = (\langle \hat{S}_i^x \hat{S}_{j}^x \rangle + \langle \hat{S}_i^y \hat{S}_{j}^y\rangle)/2$ measures transverse ($\perp$) spin correlations as the mean of the correlations in $x$ and $y$ directions.
The latter term takes longitudinal spin correlations into account, where $P^{\uparrow, \uparrow (\downarrow, \downarrow)}_{i,j}$ is the joint probability to find spin-$\up$ ($\down$) atoms at positions $i$ and $j$.

Similar to the experiments reported in~\cite{Fukuhara:2013a, Fukuhara:2013b}, we started with the preparation of a two-dimensional quantum-degenerate gas of typically $170$ $^{87}$Rb atoms in the $\down \equiv \ket{F=1,m_{F}=-1}$ state trapped in a single anti-node of a vertical ($z$-axis) optical lattice with a depth of $V_z=20\,E_r$, where $E_{r}=h^2/(8m a_{\rm lat}^2)$ is the recoil energy with lattice spacing $a_{\rm lat}=532\,$nm and atomic mass $m$.
To prepare the system for the local spin flip, we adiabatically ramped up two horizontal lattices to $V_{x,y}=40\,E_r$, which drove the gas deep into the Mott insulating phase~\cite{Sherson:2010}.
At this stage, we extracted the temperature $T = 0.08(3)\,U/k_{\rm B}$ from the density and distribution of holes~\cite{Sherson:2010}.
Notably, we improved our minimal temperature compared to earlier experiments by almost a factor of two~\cite{Fukuhara:2013a} resulting in a lower probability of $0.032(6)$ per site to be empty (in the central region of interest of nine sites).
These empty sites might be either non- or doubly-occupied sites, which we cannot distinguish in our imaging procedure due to parity projection~\cite{Sherson:2010}.
A single line of atoms (in spatial $y$-direction) in the center of the Mott insulator was then transferred to the $\up \equiv \ket{F=2,m_F=-2}$ state using our single site addressing technique~\cite{Weitenberg:2011,Fukuhara:2013a,Fukuhara:2013b}, thereby deterministically creating an impurity spin in each chain.
The wavelength and polarization of the line-shaped addressing beam was adjusted such that the $\down$ state was nearly unaffected while atoms in the $\up$ state experienced an attractive potential.
Next we decreased the lattice along the $x$-direction to $V_x=10\,E_r$ within $50\,$ms while keeping the addressing beam on.
The one-dimensional impurity dynamics with $J_{ex}/ \hbar=2\pi \times 10\,$Hz, $J/\hbar=2\pi \times 39\,$Hz and $U/\hbar=2\pi \times 800\,$Hz was then initiated by switching off the addressing beam in $1\,$ms.
Here, $J_{ex}$ was directly extracted from the impurity dynamics~\cite{Fukuhara:2013a, Fukuhara:2013b}, while $J$ and $U$ were calculated for the given lattice depths.
Corrections due to density-induced tunneling resulted in an increase of $J_{ex}$ of about $20\%$ compared to $4J^2/U$~\cite{Jurgensen:2014a}.

After a variable evolution time $t$, we froze the atomic distribution by increasing the depth of all lattices.
Selective imaging of the $\up$ state was performed by inverting the spin population using a global microwave sweep followed by a push out of the $\ket{2,-2}$ state on the cycling transition and subsequent site-resolved fluorescence imaging of the remaining atoms~\cite{Sherson:2010}.
The probability $P^1_{i \, \parallel}$ to find one atom at site $i$ after this longitudinal ($\parallel$)  measurement corresponds to the probability for the atom to be in the $\up$ state.
Clear interference fringes with high visibility were observed in these measurements (see Fig.~1b).
In order to measure the analogous quantity $P^1_{i \, \perp}$ in the transverse basis, we added a global $\pi/2$ rotation before the spin selective imaging.
Spatially homogeneous magnetic field fluctuations randomize the transverse phase within less than $1\,$ms, resulting in rotational symmetry around the longitudinal spin axis.
Consequently, the transverse spin distribution is uniform across the chain with equal probabilities for the $\up$ and $\down$ states (Fig.~1b).
We ensured that the magnetic field homogeneity was better than $50\,$mG/cm ($0.2\,$Hz/$a_{\rm lat}$)~\cite{Hild:2014} such that transverse spin correlations are preserved over experimental timescales of $100\,$ms.
Each data point presented in this paper is extracted from typically $800$ ($1000$) individual realizations of the spin chain in the longitudinal (transverse) case (with the exception of the data shown in Fig.~4, where these numbers are five times lower).

\begin{figure}[t]
\centering
\includegraphics[]{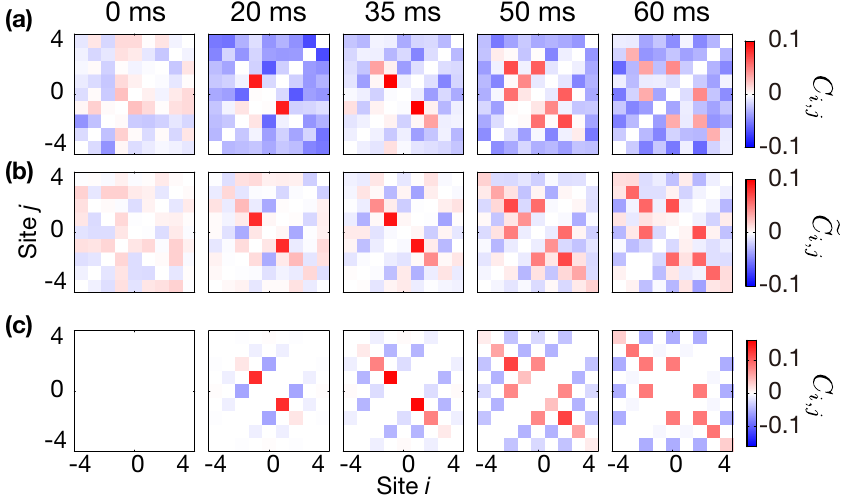}
\caption{Transverse correlations.
(a), (b) Experimental data of the correlations $C_{i,j}$ and $\widetilde{C}_{i,j}$ for different evolution times.
The strongest signal corresponds to the outward propagating correlations between the sites $\pm i$ symmetrically located around the initial position (on the upper left to lower right diagonal).
(c) Theoretical prediction for an ideal spin chain.
Remarkable qualitative agreement between theory and experiment is visible in the spatial structure of the correlations, but the amplitude of the experimental signal is reduced (different color scales).}
\label{fig:transverse_spin}
\end{figure}

The spatially resolved measurement of transverse correlations $C_{i,j}$ is the crucial step towards the detection of entanglement dynamics.
Without defects in the spin chain, the operator $\hat{S}^\perp_i$ directly relates to our experimental observable $P^1_{i \, \perp}=\langle \hat{S}^\perp_i\rangle+\frac{1}{2}$.
Hence, the transverse correlations $C_{i,j}=P_{i,j \, \perp}^{11}-\frac{1}{4}$ are given by the joint probability $P_{i,j \, \alpha}^{11}$ to find one atom at site $i$ and one at site $j$ in the transverse ($\alpha=\perp$) measurement.
Imperfections will always decrease these detected correlations such that $C_{i,j}$ provides a lower bound for them, even in an environment of onsite atom number fluctuations~\cite{Mazza:2015}.
Figure~2a shows the measured transverse correlations together with the theoretical prediction for the ideal XX-spin chain.
A strong positive signal appears between sites $+1$ and $-1$ after an evolution time of $20$\,ms ($1.26 \hbar/J_{ex}$) and subsequently these correlations spread further outwards.
However, compared to the ideal case a trend toward negative values is visible even between far separated sites that should be uncorrelated given the short evolution times.
A possible explanation for this lies in the non-perfect initial Mott insulators resulting in $P_{i\,\perp}^{1} < 0.5$, which biases the measured $C_{i,j}$.
This bias is removed in the modified transverse correlation $\widetilde{C}_{i,j}$, defined as $\widetilde{C}_{i,j}=P_{i,j\, \perp}^{11}-P_{i\,\perp}^{1} P_{j\,\perp}^{1}$~\cite{Mazza:2015}.
In Fig.~2b, we show the measured $\widetilde{C}_{i,j}$, now in remarkable agreement with the theoretical prediction, except for the smaller amplitude of the measured correlation signal.

We now combine longitudinal and transverse correlation measurements to detect spin-entanglement in the system.
This is achieved using a lower bound for the concurrence in the spin-$1/2$ degree of freedom and also for the entanglement of particles~\cite{Wiseman:2003, Schuch:2004, Dowling:2006} obtained from~\cite{Mazza:2015}
\begin{equation}
  \mathscr{C}_{i,j} = 2 ( 2C_{i,j} - \sqrt{P^{11}_{i,j\, \parallel} P^{00}_{i,j\, \parallel}} ).
  \label{eq:concurrence}
\end{equation}
Here, $P^{00}_{i,j\, \parallel}$ is the joint probability of finding zero atoms on sites $i$ and $j$ in the longitudinal measurement.
It has been shown in Ref.~\cite{Mazza:2015} that $\mathscr{C}_{i,j}$ is a valid lower bound for the concurrence even in the case of fluctuating particle numbers as long as the maximum on-site occupation number does not exceed two.
This requirement is fulfilled in our experiments where the total atom number is tuned to yield a unity filled Mott insulator in the center of the trap.
Assuming the worst case scenario, that the observed hole probability of $0.032(6)$ is only due to doubly occupied sites and an exponentially decreasing occupation of higher excited states we expect a probability for triply occupied states of $10^{-3}$.
With the weak additional assumptions of vanishing correlations between the site occupation numbers and between all degrees of freedom in the doubly occupied sector, a more efficient bound $\widetilde{\mathscr{C}}_{i,j}$ on the concurrence  can be obtained by replacing $C_{i,j}$ by the modified transverse correlations $\widetilde{C}_{i,j}$ in Eq.~(\ref{eq:concurrence})~\cite{Mazza:2015}. It is reasonable to assume that these conditions are fulfilled in the experiment for non nearest-neighbor sites and given the very low probability for double occupation.

The results of the concurrence measurements are shown in Fig.~3a for pairs of sites symmetric around the initial impurity position.
They reveal a buildup of entanglement in the spin chain leading to a peaking concurrence $\mathscr{C}_{-1,+1}=0.24 (6)$ between the sites $\pm1\,a_\mathrm{lat}$ away from the center after $35\,$ms.
For longer times, the concurrence peaks at larger distances showing an outward propagating entanglement wavefront.
Using the bound $\widetilde{\mathscr{C}}_{i,j}$, we find finite entanglement up to distances of six lattice sites.
Note that the bound for the concurrence is expected to be especially efficient for pairs of sites located symmetrically around the initial position, which is consistent with our observations (cf. Fig.~2).
The observed concurrence closely resembles the transverse correlations that are shown in Fig.~3b for comparison.
Its amplitude is only slightly decreased due to our finite fidelity in the preparation of the initial spin-impurity.
Comparing the measured transverse correlations $\widetilde{C}_{i,j}$ quantitatively to the expectation for a perfect chain, we find good agreement after a constant down scaling of the theoretical correlation amplitude by $0.6$.
Such an effect has indeed been predicted in an environment of holes and doubly occupied sites~\cite{Mazza:2015}.
An decreasing entanglement signal with increasing site separation is also expected in the ideal case due to the dispersion of the single spin-impurity.

\begin{figure}[t]
\centering
\includegraphics[]{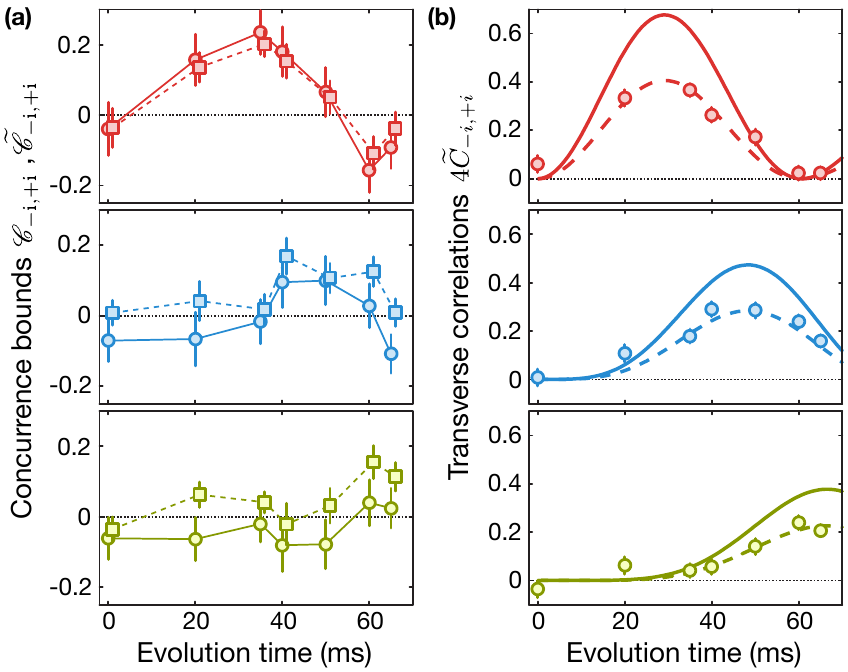}
\caption{Propagation of an entanglement wave.
(a) Experimental lower bounds for the concurrence $\mathscr{C}_{-i,+i}$ (circles with solid line) and the more efficient bound $\widetilde{ \mathscr{C}}_{-i,+i}$ (squares with dashed line) between the sites $\pm i$ versus time.
From top to bottom $i=1$ (red), $2$ (blue), $3$ (green) (b) Transverse spin correlations $4 \widetilde{C}_{-i,+i}$ for the same sites.
Circles are the experimental data and the solid lines show the predictions for the defect-free ideal case, where $4\widetilde{C}_{-i,+i}=\widetilde{ \mathscr{C}}_{-i,+i}=\mathscr{C}_{-i,+i}$. The dashed lines are the ideal predictions scaled by $0.6$. Error bars indicate the standard error of the mean (s.e.m.)}
\label{fig:concurrence}
\end{figure}

The natural question arising from these results is the impact of defects in the chain on coherence and entanglement in the spin sector. To address this question, we refined our measurement technique to simultaneously detect the local occupation number in both spin states, which also directly gives access to hole defects in the system.
Our technique is based on an in-situ Stern-Gerlach like measurement~\cite{Gerlach:1922}, which spatially separates the $\up$ and $\down$ spins prior to the detection~\cite{Mazza:2015}.
Since imaging of the atoms is restricted to a single plane, this required the preparation of single isolated one-dimensional (1d) systems such that the Stern-Gerlach separation can be done transversally.
The experimental sequence closely followed the procedure described earlier, but we additionally use the local addressing system to remove all atoms but those in a single 1d tube prior to the preparation of the spin-impurity.
After the spin evolution, we switched off the lattice in $y$-direction (transverse to the chain) and applied a magnetic field gradient along the same direction.
Due to the different magnetic moments of the $\up$ and $\down$ states, the two components were spatially separated.
Next, we switched the lattice in $y$-direction back on in $75\,$ms to $10\,E_r$.
Finally, we used our standard spin-insensitive fluorescence detection to observe the atoms locally.
In the obtained image, the position along $x$ defines the position in the spin chain and the position in $y$-direction identifies the spin state (Fig.~4a).
This measurement is very challenging, both, due to the increased complexity of the preparation and due to the lower statistics; we measure only a single chain per experimental run as compared to nine chains in the simpler standard protocol.
Therefore, we limited ourselves to a fixed evolution time of $35\,$ms, the setting for which the concurrence peaked between the $\pm1$ sites.

\begin{figure}[t]
\centering
\includegraphics{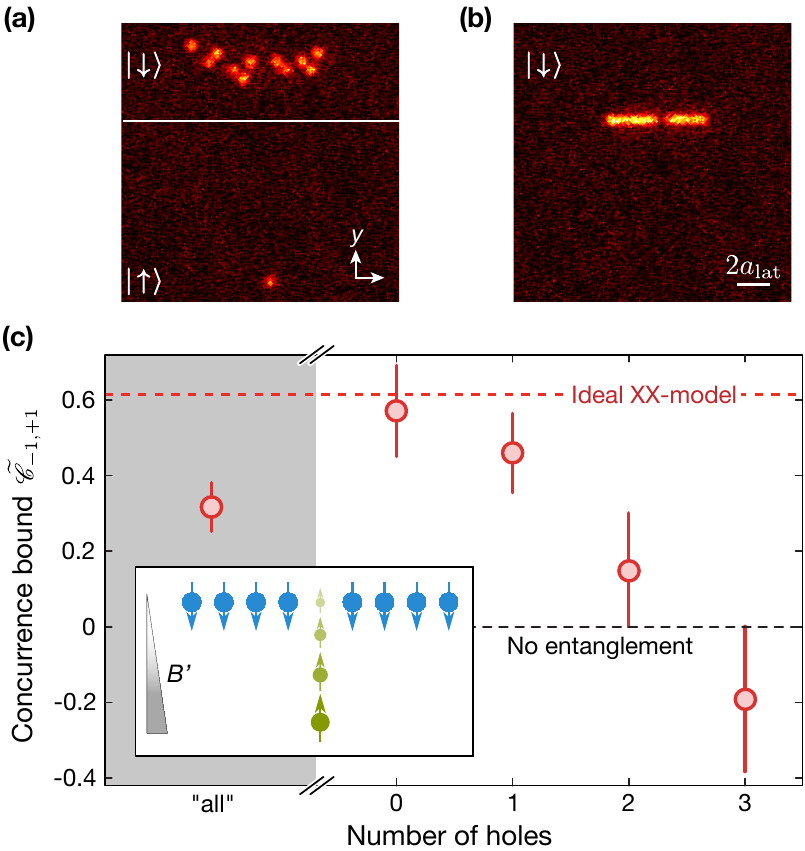}
\caption{Impact of defects on spin-entanglement.
Comparison of images of single spin chains taken with in-situ Stern-Gerlach imaging of both spin states (a) to the standard spin-resolved imaging in which one spin state was removed before detection (b). In the former case the magnetic field gradient was in vertical direction, pushing the \mbox{$\up$-spins} down and the \mbox{$\down$-spins} up from the initial position (white line).
(c) Lower concurrence bound $\widetilde{ \mathscr{C}}_{-1,+1}$ after $35\,$ms evolution for the full dataset (point in the left gray part) and for subsamples postselected to $0$, $1$, $2$ and $3$ holes in the chain.
The red dashed line indicates the value expected for the perfect XX-spin chain, the black dashed line is the boundary for the presence of entanglement.
The inset schematically shows the Stern-Gerlach separation of the two spin states in the magnetic field gradient $B'$ (depicted by the left triangle).
Error bars $1\,$s.e.m.
}
\label{fig:single_1d}
\end{figure}

The results for the concurrence extracted from the spin-resolved measurements are shown in Fig.~4c.
For the analysis, we discarded all pictures with more than one atom per $y$-tube (approx. $25\%$), as those were predominantly caused by imperfect preparation of the single 1d system.
After this, the concurrence signal $\widetilde{\mathscr{C}}_{-1,+1}$ agrees with the previously measured one within experimental uncertainty -- an indication that double occupancies had little impact on the measurements reported above.
Next, we further postselected the data to a fixed number of holes (between zero and three) within the central nine sites.
For the zero hole data set, our measured concurrence indeed matches the expectation for an ideal XX-spin chain while it rapidly decreases for increasing hole number.
For three holes, i.e. a hole density of $30\%$, no entanglement is detected any more showing that defects critically affect the coherence in the spin sector.

In conclusion, we have experimentally measured bipartite entanglement in the spin degree of freedom between two sites of an optical lattice filled with ultracold atoms.
Using a novel detection method we have shown that atom number defects are critical for the transverse coherence and hence for entanglement in the spin sector.
This method also gives access to inter-spin correlations, an important quantity to characterize two-component bosons in the superfluid regime~\cite{Fukuhara:2013a}.
Our measurements pave the way toward in-depth studies of entanglement in quantum many-body entanglement systems~\cite{Amico:2008} and mark the first steps toward controlled entanglement transfer across spin wires~\cite{Bose:2003, Subrahmanyam:2004, Bose:2007}.

\begin{acknowledgments}
  We acknowledge fruitful discussions with G. Giedke, L. Mazza, D. Rossini, and R. Fazio and funding by the MPG, EU (UQUAM) and the Körber Foundation.
\end{acknowledgments}

\bibliography{References}

\end{document}